\documentclass[useAMS,usenatbib,usegraphicx]{mn2e}
\usepackage{ gensymb }

\usepackage{graphicx}
\usepackage[fleqn]{amsmath}
\usepackage{amssymb}
\usepackage{bm}

\newcommand{\dd}{{\rm d}}

\title[Impact of neutrinos on the MRI]
{Neutrino viscosity and drag: impact on the magnetorotational instability in protoneutron stars}

\author[Guilet et al.]{J\'er\^ome Guilet$^{1,2}$, Ewald M\"uller$^{1}$ \& Hans-Thomas Janka$^{1}$ \\
$^1$ Max-Planck-Institut f\"ur Astrophysik, Karl-Schwarzschild-Str. 1, D-85748 Garching, Germany \\ 
$^2$ Max Planck/Princeton Center for Plasma Physics}

\begin{document}

\maketitle

\label{firstpage}

\begin{abstract}
The magnetorotational instability (MRI) is a promising mechanism to amplify the magnetic field in fast rotating protoneutron stars. The diffusion of neutrinos trapped in the PNS induces a  transport of momentum, which can be modelled as a viscosity on length-scales longer than the neutrino mean free path. This neutrino-viscosity can slow down the growth of MRI modes to such an extent that a minimum initial magnetic field strength of $\gtrsim 10^{12}\, {\rm G}$ is needed for the MRI to grow on a sufficiently short time-scale to potentially affect the explosion. It is uncertain whether the magnetic field of fast rotating progenitor cores is strong enough to yield such an initial magnetic  field in PNS. At MRI wavelengths shorter than the neutrino mean free path, on the other hand, neutrino radiation does not act as a viscosity but rather induces a drag on the velocity with a damping rate independent of the wavelength. We perform a linear analysis of the MRI in this regime, and apply our analytical results to the PNS structure from a one-dimensional numerical simulation. We show that in the outer layers of the PNS, the MRI can grow from weak magnetic fields at wavelengths shorter than the neutrino mean free path, while deeper in the PNS MRI growth takes place in the viscous regime and requires a minimum magnetic field strength. 
\end{abstract}

\begin{keywords}
MHD -- stars: neutron -- stars: rotation -- supernovae: general
\end{keywords}

%%%%%%%%%%%%%%%%%%%%%%%%%%%%%%%%%%%%%%%%%%%%%%%%%%%%%%%%%%%%%%%%%%%%%%%%%%%%%%%%%%
\section{Introduction}
The explosion mechanism of core collapse supernovae and in particular the role played by rotation and magnetic fields is still uncertain. The neutrino driven mechanism aided by multidimensional hydrodynamical instabilities may be responsible for explosions with normal energies of $10^{50}-10^{51}\,{\rm erg}$, though robust explosions with sufficient energy have yet to be demonstrated by three-dimensional numerical simulations including all relevant physics \citep[e.g.][]{hanke13,mezzacappa14}. A small fraction of core collapse supernovae, however, have much larger explosion energies of $\sim 10^{52}\,{\rm erg}$ \citep["hypernovae" or type Ic BL, e.g.][]{drout11}, which most likely require an additional energy reservoir beyond neutrinos. Scenarios relying on a combination of fast rotation and strong magnetic fields may be good candidates to explain such extreme explosions. The rotation energy contained in a neutron star rotating with a period of one millisecond (near break-up velocity) is indeed a sufficient energy reservoir, which could be efficiently tapped if strong magnetic fields of the order of $\sim 10^{15}\,{\rm G}$ are present. Axisymmetric simulations assuming both a strong poloidal magnetic field and fast differential rotation have for example demonstrated the possibility of magnetorotational explosions \citep{leblanc70,bisnovatyi-Kogan76,mueller79,symbalisty84,moiseenko06,shibata06,burrows07b,dessart08,takiwaki09,takiwaki11}, although it remains to be demonstrated whether the explosion energy can reach that of hypernova-like explosions. Such magnetorotational explosions are furthermore a potential site for the production of r-process elements \citep{winteler12}. Note, however, that the 3D dynamics of magnetorotational explosions needs to be explored further, since \citet{mosta14} showed that non-axisymmetric instabilities can disrupt the jet before it can launch an explosion.

Another way by which fast rotation and strong magnetic fields could impact supernovae explosions is through the delayed injection of energy due to the spin down of a fast rotating, highly magnetized neutron star \citep{kasen10,woosley10}, which has been invoked as an explanation of some superluminous supernovae like SN 2008 bi \citep{dessart12,nicholl13,inserra13}. The birth of such "millisecond magnetars" is furthermore a potential central engine for long gamma-ray bursts \citep[e.g.][]{duncan92,metzger11}.

The very fast rotation needed by the above scenarios (magnetorotational explosion, millisecond magnetars) may not be present in the core of stars following standard stellar evolution, as \citet{heger05} showed that magnetic torques can slow down the core rotation efficiently. However, \citet{yoon05} and \citet{woosley06} showed that the fastest rotating stars could follow a chemically homogeneous evolution, in which the core can retain enough angular momentum to form a neutron star rotating with milliseconds period. 

The second crucial ingredient for powerful magnetorotational energy release is an extremely strong, large-scale poloidal magnetic field of the order of $10^{15}\,{\rm G}$. The presence of such a strong magnetic field in some protoneutron stars (PNS) is suggested by the observation of the most strongly magnetized neutron stars called magnetars \citep[and references therein]{woods06}. The origin of this strong magnetic field remains, however, uncertain. One hypothesis is a fossil field origin in which the magnetic flux is inherited from the progenitor star \citep{ferrario06}, but it is not clear whether this can explain the population of magnetars. Given the uncertainty of having a sufficiently strong magnetic field in the iron core, intense research has been undertaken on physical mechanisms that could amplify the magnetic field during core collapse (in addition to the compression due to magnetic flux conservation during the collapse). In non-rotating progenitors, the standing accretion shock instability and convection have been invoked as a source of turbulence giving rise to a small-scale dynamo \citep{endeve10,endeve12,obergaulinger11,obergaulinger14}. An Alfv\'en surface has also been shown to be a potential site of Alfv\'en wave and therefore magnetic field amplification \citep{guilet11}. But the most promising mechanisms rely on the fast rotation to drive a dynamo in the convective region of the PNS \citep{thompson93}, or through the magnetorotational instability  \citep[hereafter MRI; e.g.][]{balbus91,akiyama03}.   

Since the suggestion by \citet{akiyama03} that the MRI could play an important role in core collapse supernovae, the MRI has been the subject of a number of studies in the context of supernovae with the use of linear analysis \citep{masada06,masada07}, local (or "semi-global") numerical simulations representing a small portion of the PNS \citep{obergaulinger09,masada12}, and two-dimensional global numerical simulations \citep{sawai13,sawai14}. \citet{masada07} showed that neutrino radiation in the diffusive regime has several effects on the MRI. On the one hand, neutrino thermal and lepton number diffusion alleviates the stabilizing effect of entropy and lepton number gradients in stably stratified regions of the PNS. On the other hand, neutrino viscosity slows down MRI growth if the initial magnetic field is weaker than a critical strength, which \citet{masada12} estimated to be $\sim 3.5\times10^{12}\,{\rm G}$. As a consequence of neutrino viscosity, a minimum magnetic field is needed for the MRI to grow on a sufficiently short time-scale to affect the explosion \citep{masada12}. In this paper, we revisit this issue by computing the neutrino viscosity for the conditions given by the output of a one-dimensional numerical simulation. We find that the effect of neutrino viscosity is even more pronounced: MRI growth is slowed down if the magnetic field is weaker than $10^{13}-10^{14}\,{\rm G}$ (depending on the rotation rate), and becomes too slow to affect the explosion below a minimum strength of $\sim 10^{12}\,{\rm G}$ (see Section~\ref{sec:viscous_MRI}). This may be a problem for the MRI since the initial magnetic field in the PNS is extremely uncertain and could well be below this minimum strength.

The description of the effect of neutrinos by a viscosity is however valid only at length-scales longer than the neutrino mean free path. We will show that, in the outer parts of the PNS, the viscous prescription is not self-consistent because the MRI grows at wavelengths shorter than the neutrino mean free path. We therefore provide the first description of the effect of neutrinos on the growth of the MRI that is valid at wavelengths shorter than the neutrino mean free path (Section~\ref{sec:drag_MRI}). This allows us to show that in the outer parts of the PNS, the MRI can grow from initially very weak magnetic fields.

The paper is organized as follows. In Section~\ref{sec:PNS_model}, we describe the numerical model of PNS structure. In Section~\ref{sec:regimes_MRI}, we analyse the different regimes in which the MRI can grow: analytical predictions are obtained, which are then applied to the PNS model. In Section~\ref{sec:conclusion}, we discuss and conclude on the relevant regime of MRI growth as a function of radius in the PNS and magnetic field strength.

%%%%%%%%%%%%%%%%%%%%%%%%%%%%%%%%%%%%%%%%%%%%%%%%%%%%%%%%%%%%%%%%%%%%%%%%%%%%%%%%%%
\section{PNS model}
	\label{sec:PNS_model}
\begin{figure}
\centering
 \includegraphics[width=\columnwidth]{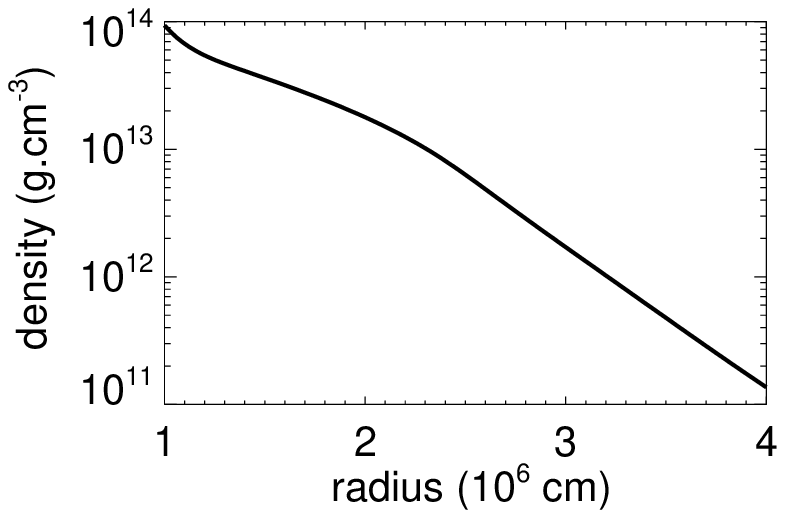}
  \includegraphics[width=\columnwidth]{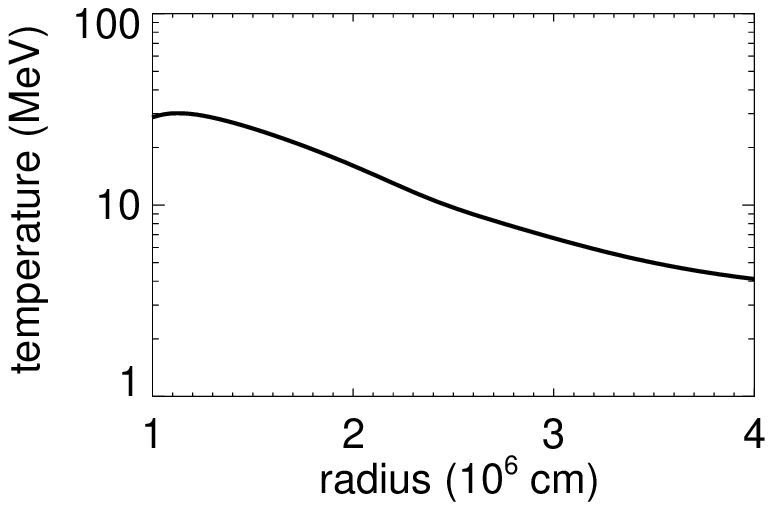}
  \includegraphics[width=\columnwidth]{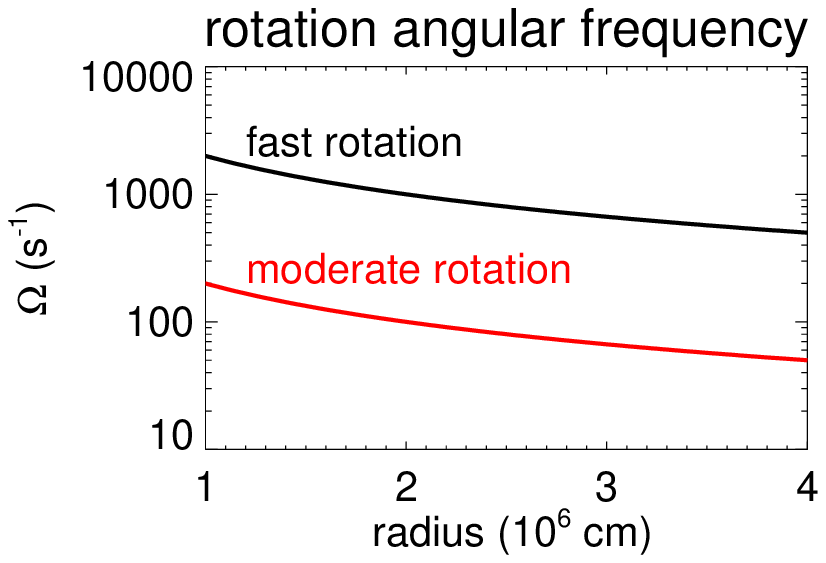}
 \caption{Structure of the PNS at time $t=170\,{\rm ms}$ after bounce. Upper panel: Radial profile of the density. Middle panel: Radial profile of the temperature.  Lower panel: radial profiles of rotation angular frequency defined in equations~(\ref{eq:rotation_profile1})-(\ref{eq:rotation_profile2}), with parameters $\Omega_0=2000\,{\rm s^{-1}}$ (black line) or $\Omega_0=200\,{\rm s^{-1}}$ (red line), $q=1$ and $r_0=10\,{\rm km}$.}
             \label{fig:PNS_structure}%
\end{figure}

In order to estimate physical quantities relevant for the growth of the MRI, we use the result of a one-dimensional numerical simulation as a typical structure of the PNS. The calculations were performed with the code Prometheus-Vertex, which combines the hydrodynamics solver Prometheus \citep{fryxell89} with the neutrino transport module Vertex \citep{ramp02}. Vertex solves the energy-dependent moment equations with the use of a variable Eddington factor closure, and including an up-to-date set of neutrino interaction rates \citep[e.g.][]{mueller12b}. General relativistic corrections are taken into account by means of an effective gravitational potential \citep{marek06}. The model considered hereafter simulates the evolution of the $11.2\,M_\odot$ progenitor of \citet{woosley02}, using the high-density equation of state of \citet{lattimer91} with a nuclear incompressibility of $K = 220\,{\rm MeV}$. Most of the results presented in this paper were obtained using a single time frame at $t=170\,{\rm ms}$ after bounce. The radial profiles of density (upper panel) and temperature (middle panel) at this (arbitrary) reference time are shown in Fig.~\ref{fig:PNS_structure}. The structure of the PNS evolves in time due to its contraction and energy and lepton emission. We have therefore performed the same analysis at other times between $t=50\,{\rm ms}$ and $t=800\,{\rm ms}$ after bounce, obtaining very similar results: the main difference is the radius of the PNS (which decreases from $70\,{\rm km}$ to about $20\,{\rm km}$), as will be discussed in Section~\ref{sec:conclusion}.

As a simplified representation of the rotation profiles obtained by \citet{ott06}, we further assume that the equatorial plane of the PNS is rotating with the following rotation profile\footnote{since we restrict our analysis of MRI growth to the equatorial plane, we do not need to specify the angular dependence of the rotation frequency}
\begin{eqnarray}
\Omega = & \Omega_0, & {\rm if}\,\,\,  r < r_0   \label{eq:rotation_profile1} \\
\Omega = & \Omega_0\left(\frac{r}{r_0}\right)^{-q}, & {\rm if}\,\,\,  r > r_0    \label{eq:rotation_profile2}
\end{eqnarray}
where $\Omega_0$ is the angular frequency of the inner solidly rotating core (at $r<r_0$), and $q$ is the power law index of the rotation profile assumed outside this inner core. Note that the MRI can grow only in regions where the angular frequency is decreasing outward, such that we only consider the outer region ($r>r_0$) in the following analysis. As typical parameters, we have chosen $q=1$, $r_0=10\,{\rm km}$ and two different values of the angular frequency: $\Omega_0 = 2000\,{\rm s^{-1}}$ (fast rotation, black line in the lower panel of Fig.~\ref{fig:PNS_structure}) or $\Omega_0 = 200\,{\rm s^{-1}}$ (moderate rotation, red curve in the lower panel of Fig.~\ref{fig:PNS_structure}). Assuming shellular rotation, the total angular momentum contained in the PNS rotating with these angular frequency profiles would be: $L = 4.7\times 10^{48}\,{\rm g\,cm^2\,s^{-1}}$ (fast rotation), and $L = 4.7\times 10^{47}\,{\rm g\,cm^2\,s^{-1}}$ (moderate rotation).  After cooling and contraction to a neutron star with a radius of  $12\, {\rm km}$ and a moment of inertia of $I=1.5\times 10^{45}\,{\rm g\,cm^2}$, the neutron star would be rotating with a period (assuming angular momentum conservation): $P=2\,{\rm ms}$ (fast rotation) and $P=20\,{\rm ms}$ (moderate rotation). The fast rotation is comparable to the angular momentum of progenitors following the chemically homogeneous evolution \citep{yoon05,woosley06}, and may be thought of as representative of the scenario of millisecond magnetar formation (if magnetic field amplification to magnetar strength is indeed achieved), which is one of the central engines considered for gamma-ray bursts and hypernovae explosions. The moderately rotating model, on the other hand, is comparable to the progenitors of \citet{heger05} and is relevant to less extreme supernovae where rotation and magnetic field amplification may still play an important role.

Note that the PNS structure is taken from a one-dimensional numerical simulation of a non-rotating progenitor. Assuming a rotation profile is therefore not self-consistent, but should give the right order of magnitude as long as the rotation is not too extreme. This will be discussed in Section~\ref{sec:conclusion}.

%%%%%%%%%%%%%%%%%%%%%%%%%%%%%%%%%%%%%%%%%%%%%%%%%%%%%%%%%%%%%%%%%%%%%%%%%%%%%%%%%%
\section{Different regimes of the MRI}
	\label{sec:regimes_MRI}

In this section, we obtain analytical estimates for the effect of neutrino radiation on the MRI growth in different regimes, corresponding to optically thick or optically thin neutrino transport at the MRI wavelength. In Section~\ref{sec:ideal_MRI}, we recall classical results on the linear growth of the MRI in ideal MHD (neglecting the effects of neutrino radiation). In Section~\ref{sec:viscous_MRI}, we study the effect of neutrino viscosity on the growth of the MRI, which applies when the wavelength of the MRI exceeds the neutrino mean free path. In Section~\ref{sec:drag_MRI}, we then consider the growth of the MRI at scales shorter than the neutrino mean free path.

In order to highlight these different regimes in a simple way, we have chosen to make a number of simplifying assumptions. First, the growth rates are obtained from the results of a local WKB analysis, which applies when the wavelength is much shorter than the scale of the gradients (i.e. typically the density scaleheight). Secondly, we assume the initial magnetic field to be purely vertical, which is the most favourable configuration for MRI growth. The fastest growing MRI modes are then axisymmetric, with a purely vertical wave vector \citep[e.g.][]{balbus91}, and we therefore make this assumption in the linear dispersion relations presented below. Thirdly, we assume that the gas is incompressible, i.e. we assume that the sound speed is much larger than the velocities and Alfv\'en speed (which should be reasonably well justified at least in the linear phase of the MRI) and neglect buoyancy effects due to the presence of entropy and composition gradients. Finally, we neglect the resistivity in all the scaling relations used to estimate the growth rate and wavelength of the MRI in the conditions prevailing inside the PNS. This assumption is justified by the very small value of the resistivity compared to the neutrino viscosity \citep[by a factor of about $10^{13}$, e.g.][]{thompson93,masada06}. For completeness, however, resistive effects are retained in the dispersion relations presented in Sections~\ref{sec:viscous_MRI} and \ref{sec:drag_MRI}.

%%%%%%%%%%%%%%%%%%%%%%%%%%%%%%%%%%%%%%%%%%%%%%%%%%%%%%%%%%%%%%%%%%%%%%%%%%%%%%%%%%
\subsection{MRI in ideal MHD}
	\label{sec:ideal_MRI}
Neglecting all diffusion coefficients as well as the effect of neutrino radiation, the dispersion relation of the axisymmetric MRI modes in the presence of a vertical magnetic field is \citep[e.g.][]{balbus91}
\begin{equation}
\left(\sigma^2 +k^2v_A^2\right)^2  + \kappa^2 \left(\sigma^2 + k^2v_A^2 \right) - 4\Omega^2k^2v_A^2  = 0,
	\label{eq:dispersion_ideal}
\end{equation}
where $\sigma$ is the growth rate, $k$ is the wavenumber, $v_A \equiv B/\sqrt{4\pi\rho}$ is the Alfv\'en velocity and $\kappa$ is the epicyclic frequency defined by $\kappa^2 \equiv  \frac{1}{r^3} \frac{\dd (r^4 \Omega^2)}{\dd r}$. The analytical solution of this dispersion relation gives the growth rate and wavenumber of the fastest growing mode as
\begin{equation}
\sigma = \frac{q}{2}\Omega,
	\label{eq:sigma_ideal}
\end{equation}
\begin{equation}
k = \sqrt{q\left(1-q/4\right)} \frac{\Omega}{v_A},
	\label{eq:k_ideal}
\end{equation}
where $q \equiv - \dd \log\Omega/\dd\log r$ (consistent with the definition of the rotation profile in Section~\ref{sec:PNS_model}). The growth rate is independent of the magnetic field strength, and is extremely fast for rapid rotation
\begin{equation}
\sigma = 500\, q \left(\frac{\Omega}{1000\,{\rm s^{-1}}} \right) \,{\rm s^{-1}}.
\end{equation}
The wavelength on the other hand is proportional to the magnetic field strength, such that weak magnetic fields lead to very short MRI wavelength in the ideal MHD case. Assuming $q=1$ leads to the following estimate for the wavelength
\begin{equation}
\lambda = 6  \left(\frac{B}{10^{12}\,{\rm G}} \right)\left(\frac{\rho}{10^{13}\,{\rm g\,cm^{-3}}} \right)^{-1/2} \left(\frac{\Omega}{1000\,{\rm s^{-1}}} \right)^{-1} \,{\rm m}.
\end{equation}

%%%%%%%%%%%%%%%%%%%%%%%%%%%%%%%%%%%%%%%%%%%%%%%%%%%%%%%%%%%%%%%%%%%%%%%%%%%%%%%%%%
\subsection{MRI in the presence of neutrino viscosity}
	\label{sec:viscous_MRI}
\subsubsection{Neutrino viscosity in the PNS}
\begin{figure}
\centering
 \includegraphics[width=\columnwidth]{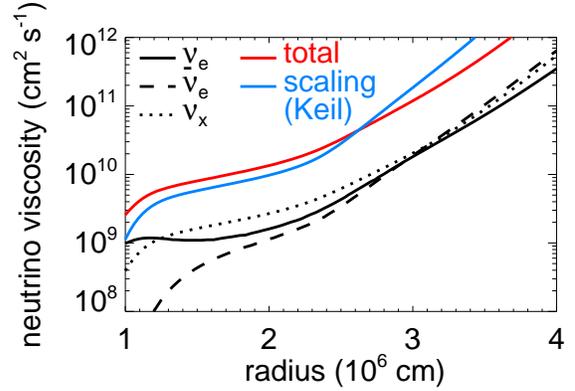}
 \caption{Radial profile of the neutrino viscosity, computed by applying equation~(\ref{eq:neutrino_viscosity}) to the outputs of the numerical simulation. The solid black line corresponds to electron neutrinos, the dashed line to electron antineutrinos, the dotted line to heavy lepton neutrinos $\nu_x$ (one representative species) and the solid red line to the total neutrino viscosity (i.e. the sum over the six neutrinos species). The blue solid line shows the approximate analytical estimate given by equation~(\ref{eq:neutrino_viscosity_approx}).}
              \label{fig:neutrino_viscosity}%
\end{figure}	

Neutrinos present in a nascent PNS can transport energy, lepton number and momentum. At length-scales much larger than the mean free path of neutrinos, their transport can be well described by diffusive processes. In this regime, the transport of momentum by neutrinos gives rise to a viscosity $\nu$  \citep{van_den_horn84,burrows88,thompson93}. It can be expressed as
\begin{equation}
\nu = \frac{2}{15} \frac{e_\nu \langle l_\nu \rangle}{\rho c}, 
	\label{eq:neutrino_viscosity}
\end{equation}
where c is the speed of light, $e_\nu$ is the neutrino energy density, and $ \langle l_\nu \rangle$ is the neutrino mean free path averaged over energy following
\begin{equation}
\langle l_\nu \rangle \equiv \left(\int \frac{\dd e_\nu}{\dd \epsilon} l_\nu \dd \epsilon \right)/e_\nu,
	\label{eq:energy_average_viscosity}
\end{equation}
with $\epsilon$ the neutrino energy. Note that the numerical factor in equation~(\ref{eq:neutrino_viscosity}) is different to that of \citet{van_den_horn84} due to our different definition of the neutrino mean free path averaged over energy. 

We compute the viscosity caused by the neutrinos of different flavors by applying equation~(\ref{eq:neutrino_viscosity}) to the output of the numerical simulation described in Section~\ref{sec:PNS_model}. The sum of the contributions from the 6 neutrino species gives the total neutrino viscosity, which varies between a few $10^{9}\,{\rm cm^2\,s^{-1}}$ near the inner boundary of the differentially rotating envelop and $10^{12}\,{\rm cm^2\,s^{-1}}$ near the neutrinosphere (Fig.~\ref{fig:neutrino_viscosity}). An approximate analytical expression for the neutrino viscosity as a function of density and temperature has been obtained by \citet{keil96} by considering six species of non-degenerate neutrinos in local thermodynamic equilibrium and assuming that the opacity comes only from scattering on to neutrons and protons in non-degenerate nuclear matter
\begin{equation}
\nu = 1.2\times 10^{10} \left(\frac{T}{10\,{\rm MeV}}\right)^2\left(\frac{\rho}{10^{13}\,{\rm g\,cm^{-3}}}\right)^{-2} \,{\rm cm^2\,s^{-1}}. 
	\label{eq:neutrino_viscosity_approx}
\end{equation}
This analytical formula is compared with the viscosity computed from the output of the numerical simulations in Fig.~\ref{fig:neutrino_viscosity}. At radii $13\,{\rm km}<r<23\,{\rm km}$, the analytical estimate reproduces the slope well and is in agreement with the numerical result within $30\%$ (the difference is due to different prescriptions for opacity, and the neglect of degeneracy for electron neutrinos and antineutrinos). At $r<13\,{\rm km}$, high-density effects such as fermion blocking and nucleon correlation effects increase the mean free path in the numerical model but are neglected in equation~(\ref{eq:neutrino_viscosity_approx}), which therefore underestimates the viscosity. At $r>23{\rm km}$, the analytical estimate increases faster than the value computed from the output of the simulation. This can be traced back to the fact that the heavy lepton neutrinos and electron antineutrinos are not perfectly in thermal equilibrium with the gas as assumed for equation~(\ref{eq:neutrino_viscosity_approx}).

It is interesting to compare this result to the literature. \citet{masada07,masada12} estimated a typical value of the neutrino viscosity $\nu = 10^{10} \,{\rm cm^2\,s^{-1}}$ at a density of $10^{12}\,{\rm g\,cm^{-3}}$. This is about 10 times smaller than what we find at the same density (at a radius of $\sim 32\,{\rm km}$): $\nu\sim 1-2\times 10^{11}\,{\rm cm^2\,s^{-1}}$. \citet{thompson05} have computed the neutrino viscosity due to different species of neutrinos (their Figure 5). The contribution from muon neutrinos varies between roughly a few times $10^{10}$ and $10^{11}\,{\rm cm^2s^{-1}}$, while our results show more variation between $\sim 10^9\,{\rm cm^2s^{-1}}$ and a few times $\sim 10^{11}\,{\rm cm^2s^{-1}}$. This difference might be due to a different PNS structure (for example due to the different time ($105\,{\rm ms}$ in their case) and equation of state). 

In the next subsection, numerical estimates will use as fiducial values representative of a radius at $20-25\,{\rm km}$ from the centre: a viscosity: $\nu = 2 \times 10^{10} \,{\rm cm^2s^{-1}}$, a density $\rho=10^{13}\,{\rm g\,cm^{-3}}$ and a rotation angular frequency $\Omega=1000\,{\rm s^{-1}}$ (fast rotation).

\subsubsection{Effect on the MRI}
\begin{figure}
\centering
 \includegraphics[width=\columnwidth]{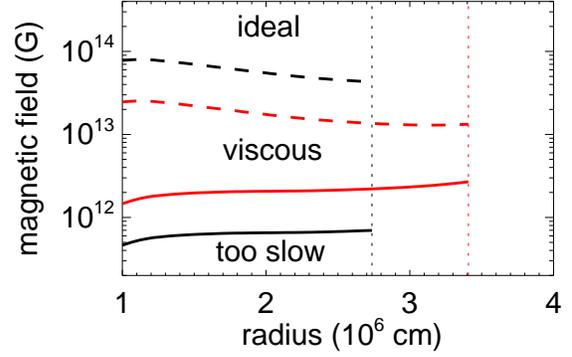}
 \caption{Radial profile of the critical magnetic field strengths that determine the regime of MRI growth. The dashed lines show the magnetic field strength $B_{\rm visc}$ (defined in equation~(\ref{eq:Bvisc})) below which viscous effects are important. The solid lines show the minimum magnetic field strength necessary for the MRI to grow at a growth rate faster than $\sigma_{\rm min} = 10\,{\rm s^{-1}}$ (computed using equation~(\ref{eq:viscous_Bmin})). The vertical dotted lines show the radius above which the viscous description breaks down because the mean free path of heavy lepton neutrinos becomes larger than the wavelength of the MRI. The two colours represent the two different normalizations of the rotation profile: fast rotation (black) and moderate rotation (red). }
              \label{fig:Bmin}%
\end{figure}

The dispersion relation of the MRI in the presence of a viscosity $\nu$ and a resistivity $\eta$ can be written as \citep{lesur07,pessah08,masada08} 
\begin{equation}
\left(\sigma_\nu\sigma_\eta +k^2v_A^2\right)^2  + \kappa^2 \left(\sigma_\eta^2 + k^2v_A^2 \right) - 4\Omega^2k^2v_A^2  = 0,
	\label{eq:dispersion_viscous}
\end{equation}
where $\sigma_\nu \equiv \sigma + k^2\nu $ and $\sigma_\eta \equiv \sigma + k^2\eta $. The effect of viscosity on the linear growth of the MRI is controlled by the viscous Elsasser number $E_\nu \equiv \frac{v_{A}^2}{\nu\Omega}$ \citep[e.g.][]{pessah08,longaretti10}. For $E_\nu < 1$, viscosity affects significantly the growth of the MRI: as a result the growth rate is decreased, and the wavelength of the most unstable mode becomes longer. Typical conditions inside the PNS lead to the following estimate of the viscous Elsasser number for fast rotation at a radius $\sim 20-25\,{\rm km}$
\begin{eqnarray}
E_\nu &\sim& 4\times 10^{-4} \left(\frac{B}{10^{12}\,{\rm G}} \right)^{2}\left(\frac{\rho}{10^{13}\,{\rm g\,cm^{-3}}} \right)^{-1} \left(\frac{\Omega}{1000\,{\rm s^{-1}}} \right)^{-1} \nonumber \\
&&\times \left(\frac{\nu}{2\times10^{10}\,{\rm cm^2\,s^{-1}}} \right)^{-1}.
\end{eqnarray}
Viscosity therefore has a large effect on the linear growth of the MRI, unless the magnetic field is initially quite strong. The critical strength of the magnetic field below which viscous effects become important (at which $E_\nu=1$) is
\begin{eqnarray}
B_{\rm visc} &=& \sqrt{4\pi\rho \nu\Omega} 	\label{eq:Bvisc} \\
&=& 5\times10^{13}\left(\frac{\rho}{10^{13}\,{\rm g\,cm^{-3}}} \right)^{1/2} \left(\frac{\nu}{2\times10^{10}\,{\rm cm^2\,s^{-1}}} \right)^{1/2} \nonumber \\
&&\times \left(\frac{\Omega}{1000\,{\rm s^{-1}}} \right)^{1/2} \,{\rm G}.
\label{eq:Bvisc2}
\end{eqnarray}
This critical magnetic field strength is shown as a function of radius in the PNS in Fig.~\ref{fig:Bmin} (dashed lines). It decreases outward but only weakly because the effect of the decrease of density and angular frequency is partly compensated by the increase in neutrino viscosity. The initial magnetic field should be quite large so that MRI growth is not much affected by viscosity :  $B \gtrsim 5\times10^{13}-10^{14}\,{\rm G}$ for fast rotation, and $B \gtrsim 1-2\times10^{13}\,{\rm G}$ for moderate rotation. Note that these values are significantly larger than the one estimated by \citet{masada12} ($\sim 3.5\times10^{12}\,{\rm G}$), which is due to the fact that the viscosity we have computed is significantly larger than the one they have estimated.

\begin{figure*}
\centering
 \includegraphics[width=2\columnwidth]{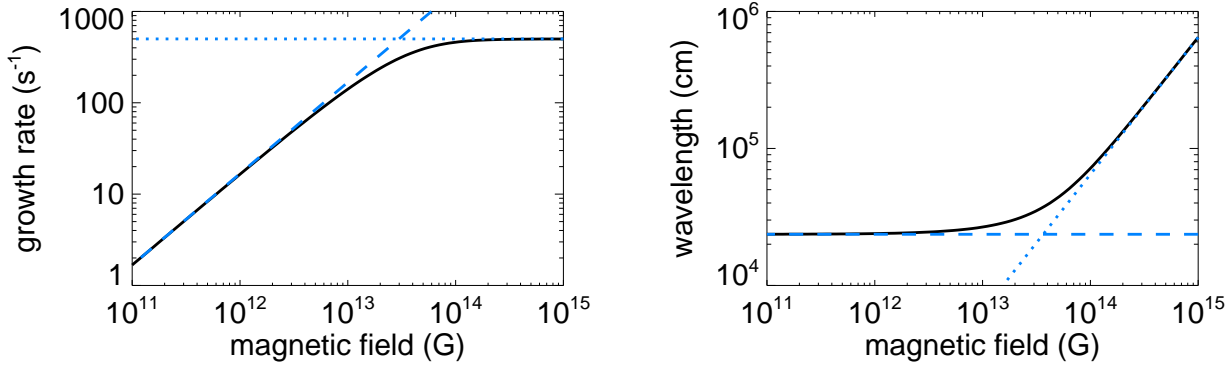}
 \caption{Growth rate (left) and wavelength (right) of the fastest growing MRI mode  in the viscous regime as a function of the magnetic field strength for the following fiducial parameters: $\nu=2\times10^{10}\,{\rm cm^2\,s^{-1}}$, $\rho=10^{13}\,{\rm g\, cm^{-3}}$ and $\Omega =1000\,{\rm s^{-1}}$. The solid black line shows the numerical solution of the dispersion relation (equation~\ref{eq:dispersion_viscous}). The dashed line shows the asymptotic behaviour in the viscous regime ($E_\nu \ll 1$), given by equations~(\ref{eq:sigma_viscous}) and (\ref{eq:k_viscous}). The dotted line shows the asymptotic behaviour in the ideal regime ($E_\nu \gg 1$), given by equations~(\ref{eq:sigma_ideal}) and (\ref{eq:k_ideal}).}
             \label{fig:viscous_MRI}%
\end{figure*}

In order to describe the MRI growth in the viscous regime, useful analytical formulae can be obtained in the asymptotic limit $E_\nu \ll 1$ (and neglecting the resistivity). The growth rate and wavenumber of the most unstable mode can then be expressed as \citep{pessah08,masada08,masada12}
\begin{equation}
\sigma = \left(\frac{qE_\nu}{\tilde\kappa} \right)^{1/2} \Omega,
	\label{eq:sigma_viscous}
\end{equation}
\begin{equation}
k =\left(\frac{\kappa}{\nu} \right)^{1/2},
	\label{eq:k_viscous}
\end{equation}
where $\tilde\kappa \equiv \kappa/\Omega = \sqrt{2(2-q)}$ is the dimensionless epicyclic frequency. In contrast to the ideal MHD case, the wavelength of the fastest growing mode is independent of the magnetic field strength (because it is set by the viscous length-scale), while the growth rate is proportional to the magnetic field strength: weak magnetic fields lead to slower growth. Using equations~(\ref{eq:sigma_viscous}) and (\ref{eq:k_viscous}), we obtain the following estimates for the growth rate and wavelength of the fastest growing MRI mode in the viscous regime
\begin{eqnarray}
\sigma &=& 17  \left(\frac{B}{10^{12}\,{\rm G}} \right)\left(\frac{\rho}{10^{13}\,{\rm g\,cm^{-3}}} \right)^{-1/2} \nonumber \\
&& \times \left(\frac{\nu}{2\times10^{10}\,{\rm cm^2\,s^{-1}}} \right)^{-1/2} \left(\frac{\Omega}{1000\,{\rm s^{-1}}} \right)^{1/2} \,{\rm s^{-1}},
	\label{eq:sigma_viscous_scaling}
\end{eqnarray} 
\begin{equation}
\lambda = 240  \left(\frac{\Omega}{1000\,{\rm s^{-1}}} \right)^{-1/2}\left(\frac{\nu}{2\times10^{10}\,{\rm cm^2\,s^{-1}}} \right)^{1/2} \,{\rm m}.
	\label{eq:wavelength_viscous}
\end{equation}
Compared to the ideal regime described in Section~\ref{sec:ideal_MRI}, the wavelength is much longer and the growth much slower for a moderate magnetic field of $10^{12}\,{\rm G}$. The growth rate and wavelength of the most unstable MRI mode are shown in Fig.~\ref{fig:viscous_MRI} as a function of magnetic field strength (and the fiducial parameters for the viscosity, density and angular frequency representative of a radius at $20-25\,{\rm km}$). The numerical solution of the dispersion relation (black solid line) is compared to the analytical solution in the asymptotic limits of ideal MRI $E_\nu \gg 1$ (dotted blue lines, Section~\ref{sec:ideal_MRI}) and viscous MRI $E_\nu \ll 1$ (dashed blue lines). The solution agrees within $20\%$ with the ideal limit for $E_\nu > 1.5$ or $B> 6\times10^{13} \,{\rm G}$ (within $10\%$ for $E_\nu > 4$ or $B> 10^{14} \,{\rm G}$), and with the viscous limit for $E_\nu < 0.05$ or $B<10^{13}\,{\rm G}$ (within $10\%$ for $E_\nu < 10^{-2}$ or $B< 5\times10^{12} \,{\rm G}$).

Fig.~\ref{fig:viscous_MRI} shows that, because of neutrino viscosity, the MRI requires a minimum magnetic field strength in order to grow fast enough to affect the explosion \citep[as was already discussed by][]{masada12}. The minimum magnetic field necessary for the MRI to grow at a minimum growth rate $\sigma_{\rm min}$ can be expressed using the viscous limit (equation~\ref{eq:sigma_viscous})
\begin{eqnarray}
B_{\rm min} &=& \left(\frac{4\pi\rho \tilde\kappa \nu}{q\Omega} \right)^{1/2}\sigma_{\rm min},  \label{eq:viscous_Bmin}	\\
&=& 6\times10^{11} \left(\frac{\sigma_{\rm min}}{10\,{\rm s^{-1}}} \right) \left(\frac{\rho}{10^{13}\,{\rm g\,cm^{-3}}} \right)^{1/2} \nonumber \\
&& \left(\frac{\nu}{2\times10^{10}\,{\rm cm^2\,s^{-1}}} \right)^{1/2} \left(\frac{\Omega}{1000\,{\rm s^{-1}}} \right)^{-1/2} {\rm G}.   \label{eq:viscous_Bmin2}
\end{eqnarray}
Fig.~\ref{fig:Bmin} shows this minimum magnetic field strength\footnote{This growth rate corresponds to an e-folding growth time of $\sigma^{-1} = 100\,{\rm ms}$ and is only a very rough estimate of the minimum growth rate needed for a significant magnetic field amplification. A significant amplification of the magnetic field by the MRI probably requires many e-folding times, and might actually take several seconds with this minimum growth rate. It is therefore possible that a dynamical effect of the MRI in the first second after shock bounce might require larger magnetic fields by a factor of a few.} as a function of radius, for $\sigma_{\rm min}=10\,{\rm s^{-1}}$. Similarly to the critical field $B_{\rm visc}$, this minimum magnetic field strength $B_{\rm min}$ depends very weakly on the radial position in the PNS (but this time slightly increasing outward due to the different scaling with $\Omega$ and the same scaling with density and viscosity). $B_{\rm min}$ is not so small ($\sim 6\times10^{11}\,{\rm G}$ for fast rotation, and $\sim 2\times10^{12}\,{\rm G}$ for moderate rotation), and we therefore conclude that neutrino viscosity sets a strong constraint on the initial magnetic field for the MRI to be able to grow on a sufficiently short time-scale. 

\begin{figure}
\centering
 \includegraphics[width=\columnwidth]{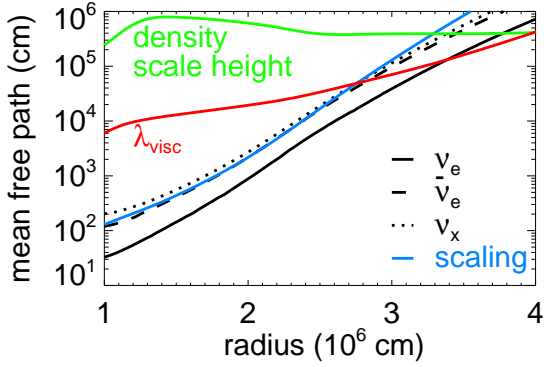}
 \caption{Radial profile of the neutrino mean free path. The solid black line corresponds to electron neutrinos, the dashed line to electron antineutrino and the dotted line to heavy lepton neutrinos $\nu_x$. The mean free paths are average over neutrino energy following equation~(\ref{eq:energy_average_viscosity}). The blue line shows the scaling $l_\nu = 10^4 (\rho/10^{13}\,{\rm g\,cm^{-3}})^{-1}(T/10\,{\rm MeV})^{-2}$. For comparison, the density scaleheight is plotted with a green line, and the wavelength of the MRI in the viscous regime in red. For the latter we use the fast rotation profile defined in equations~(\ref{eq:rotation_profile1})-(\ref{eq:rotation_profile2}) with $\Omega_0=2000\,{\rm s^{-1}}$, $q=1$ and $r_0=10\,{\rm km}$. }
             \label{fig:neutrino_mean_free_path}%
\end{figure}

This constraint, however, only applies if the MRI wavelength is longer than the mean free path of neutrinos, which may not be the case everywhere in the PNS. Fig.~\ref{fig:neutrino_mean_free_path} shows the mean free path of the different species of neutrinos as a function of radius (the mean free path is averaged over neutrino energy following equation~(\ref{eq:energy_average_viscosity})). It increases by four to five orders of magnitudes from $\langle l_\nu \rangle \sim 1\,{\rm m}$ at $r_0=10\,{\rm km}$ to $\langle l_\nu \rangle \gtrsim 10\,{\rm km}$ near the PNS surface around $40\, {\rm km}$. This dependence is approximately reproduced by the scaling $l_\nu \propto \rho^{-1}T^{-2}$ (due to the main opacity contributions being proportional to the density and the square of the neutrino energy). The wavelength of the MRI in the viscous regime is independent of magnetic field strength (equation~\ref{eq:wavelength_viscous}), and can be compared to the neutrino mean free path to check the consistency of the description (it is shown with the red curve in Fig.~\ref{fig:neutrino_mean_free_path} for the case of fast rotation). Because of the strong variation of the neutrino mean free path, the description of the effect of neutrino radiation as a viscosity is well justified deep inside the PNS, but breaks down at larger radii where the mean free path becomes longer than the MRI wavelength. This happens at radii $\sim 27\,{\rm km}$ for fast rotation and $\sim 34\,{\rm km}$ for moderate rotation, which are marked by the vertical dashed lines in Fig.~\ref{fig:Bmin}. At larger radii, the effect of neutrinos on the MRI cannot be described by a viscosity because neutrino transport begins to enter the non-diffusive regime.

%%%%%%%%%%%%%%%%%%%%%%%%%%%%%%%%%%%%%%%%%%%%%%%%%%%%%%%%%%%%%%%%%%%%%%%%%%%%%%%%%%

\subsection{MRI growth at wavelengths shorter than the neutrino mean free path}
	\label{sec:drag_MRI}
\subsubsection{Neutrino drag}

\begin{figure}
\centering
 \includegraphics[width=\columnwidth]{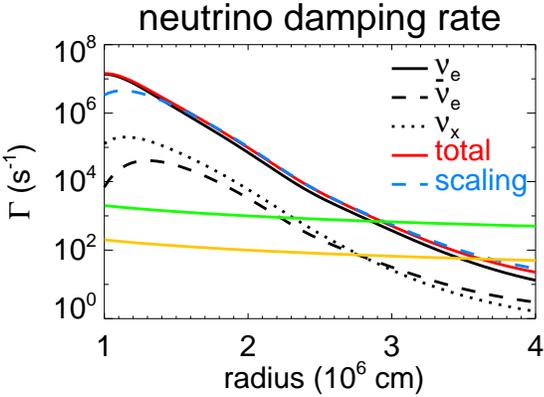}
 \caption{Radial profile of the neutrino drag damping rate computed according to equation~(\ref{eq:neutrino_drag}). The solid black line corresponds to electron neutrinos, the dashed black line to electron antineutrinos, the dotted black line to heavy lepton neutrinos $\nu_x$ (one representative species), and the solid red line to the total neutrino damping rate (i.e. the sum over the six neutrino species). The dashed blue line shows the scaling $\Gamma = 6\times 10^3\,(T/10\,{\rm MeV})^6\,{\rm s^{-1}}$. For comparison, the green and orange solid lines show the angular frequency $\Omega$ for the cases of fast rotation (green) and slow rotation (orange).}
              \label{fig:neutrino_drag}%
\end{figure}

At length-scales shorter than the neutrino mean free path, the transport of momentum by neutrinos can no longer be described as a viscous process. This momentum transport does nevertheless damp velocity fluctuations. \citet{agol98} and \citet{jedamzik98} showed that, in this regime, radiation induces a drag on the velocity field which is independent of the wavenumber of the velocity fluctuations. This drag is caused by the Doppler effect due to fluid motion with respect to the background neutrino radiation field. This creates a neutrino flux in the rest frame of the fluid, which is responsible for the drag upon absorption or scattering of the neutrinos. The neutrino drag can be represented by an acceleration $-\Gamma {\bf \delta u}$, where $\bf \delta u$ is the velocity perturbation and the damping rate $\Gamma$ is given by
\begin{equation}
\Gamma \sim \frac{e_\nu\langle\kappa_\nu\rangle}{\rho c},
	\label{eq:neutrino_drag}
\end{equation}
where $\langle \kappa_\nu \rangle$ is the neutrino opacity averaged over energy in the following way
\begin{equation}
\langle \kappa_\nu \rangle \equiv \left(\int \frac{\dd e_\nu}{\dd \epsilon} \kappa_\nu \dd \epsilon \right)/e_\nu.
	\label{eq:energy_average_drag}
\end{equation}
\citet{agol98} obtained this result (with a numerical factor of $4/3$) for photons subject to Thomson scattering, by performing a linearization of the transport equation around an isotropic and steady radiation field, using a closure model that keeps terms up to the quadrupole and neglecting higher order (note that this is a higher order approximation than the so called Eddington closure). 
\citet{jedamzik98} found a similar result for neutrinos but did not derive the relevant numerical factor, which is most likely different from the one obtained by \citet{agol98} for photons. 

Fig.~\ref{fig:neutrino_drag} shows the radial profile of the damping rate caused by the different neutrino species, obtained by evaluating equation~(\ref{eq:neutrino_drag}) for the output of the numerical simulation. The contribution from electron neutrinos is dominant compared to that of other neutrino species because of their shorter mean free path. The damping rate decreases outward by about six orders of magnitudes from $10^7\,{\rm s^{-1}}$ at $10\,{\rm km}$ to $10\,{\rm s^{-1}}$ at $40\,{\rm km}$. This can be explained by the scaling $\Gamma \propto T^6$ (shown with a blue line in Fig.~\ref{fig:neutrino_drag}), which results from the assumption of thermal equilibrium $e_\nu \propto T^4$ and $\kappa_\nu \propto \rho T^2$. \citet{thompson05} obtained a neutrino damping rate of $\Gamma \sim 50\,{\rm s^{-1}}$ at a radius of $50\,{\rm km}$ (they do not show the values at smaller radii), which is of the same order of magnitude as what we obtain at the surface of the PNS.

Note that the damping rate $\Gamma$ is related to the neutrino viscosity through $\Gamma \sim 15\nu \langle \kappa_\nu \rangle /(2\langle l_\nu \rangle) \sim \nu/\langle l_\nu\rangle^2$, i.e. it is (within a numerical factor) the rate at which fluctuations on the same scale as the neutrino mean free path would be damped by the neutrino viscosity. As a consequence the damping of fluctuations at scales shorter than the neutrino mean free path is less efficient by a factor $\sim k^2\langle l_\nu \rangle^2$ (with $k$ the wavenumber of the fluctuations) than what would be obtained by applying the viscous formalism at these scales. It is therefore conceivable that the MRI could grow on length-scales shorter than the neutrino mean free path even if the viscous formalism predicts the growth to be too slow.

\subsubsection{Growth of the MRI in the presence of neutrino drag}
	\label{sec:drag_MRI_linear}
In this section we study the growth of the MRI in the presence of neutrino drag, which is relevant to scales shorter than the neutrino mean free path. The equations of incompressible MHD in the shearing sheet approximation \citep{goldreich65} are written as
\begin{eqnarray}
\label{eq:base1}\partial_t \bm{u}+\bm{u}\cdot \bm{\nabla}
\bm{u}&=&-\frac{1}{\rho}\bm{\nabla}P+\frac{1}{\mu_0
\rho}(\bm{\nabla}\times \bm{B})\times\bm{B}\\
\nonumber
& & -2\bm{\Omega}\times\bm{u} + 2\Omega Sx\bm{e_x} - \Gamma \bm{\delta u}, \\
\label{eq:base2}\partial_t \bm{B}&=&\bm{\nabla}\times(\bm{u}\times\bm{B})+\eta\bm{\Delta{B}},\\
\label{eq:base3}\nabla \cdot \bm{u}&=&0,\\
\label{eq:base4}\nabla\cdot\bm{B}&=&0,
\end{eqnarray}
where $\bm{\Omega} \equiv \Omega\bm{e_z} $, $S\equiv q\Omega$ is the shear rate, and $\bm{e_x}$,$\bm{e_y}$,$\bm{e_z}$ are units vector in the radial, azimuthal and vertical directions (we restrict the analysis to the equatorial plane of the PNS). $\bm{\delta u}\equiv \bm{u}-\bm{u_0}$ and $\bm{\delta B} \equiv \bm{B} - \bm{B_0}$ are the velocity and magnetic field perturbations with respect to the stationary equilibrium solution $\bm{B_0}=B\bm{e_z}$, $\bm{u_0} = -q\Omega x\bm{e_y}$. The only non-standard term in these equations is the neutrino drag $-\Gamma\bm{\delta u}$ discussed in the preceding section. The velocity and magnetic field perturbations then follow the following set of equations
\begin{eqnarray}
\label{eq:pert1} \sigma \delta u_x &=& \frac{B}{\mu_0\rho}ik\delta B_x  + 2 \Omega \delta u_y  - \Gamma \delta u_x, \\
\label{eq:pert2} \sigma \delta u_y -S \delta u_x &=&   \frac{B}{\mu_0\rho}ik\delta B_y  -2 \Omega \delta u_x - \Gamma \delta u_y,\\
\label{eq:pert3} \delta u_z &=& 0, \\
\label{eq:pert4} \sigma \delta B_x&=& B ik \delta u_x - k^2\eta\delta B_x,\\
\label{eq:pert5} \sigma \delta B_y &=& B ik \delta u_y - S \delta B_x - k^2\eta\delta B_y,\\
\label{eq:pert6} \delta B_z &=&0,
\end{eqnarray}
where the perturbations are assumed to have the following time and space dependence $\delta A \propto e^{\sigma t + ikz }$, with $\sigma$ being the growth rate of the modes, and $k$ their vertical wave vector. Note that these equations are actually valid for any amplitude of the perturbations, as no linearization had to be done to obtain them (just like the channel modes of the classical MRI are non-linear solutions in the incompressible limit \citep{goodman94}). These equations can be combined to obtain the dispersion relation of the MRI in this regime
\begin{equation}
\left(\sigma_v\sigma_\eta +k^2v_A^2\right)^2  + \kappa^2 \left(\sigma_\eta^2 + k^2v_A^2 \right) - 4\Omega^2k^2v_A^2  = 0,
	\label{eq:dispersion_drag}
\end{equation}
where we have defined
\begin{equation}
\sigma_v \equiv \sigma + \Gamma.
\end{equation}
Note that the form of this equation is very similar to the dispersion relation in the viscous-resistive regime given by equation~(\ref{eq:dispersion_viscous}), the only difference being that $\sigma_\nu = \sigma + k^2\nu$ has been replaced by $\sigma_v = \sigma+ \Gamma$. As we will show, the fact that the neutrino damping rate is independent of the wavenumber makes a big difference for the wavelength and growth rate of the fastest growing mode.

\begin{figure*}
\centering
 \includegraphics[width=2\columnwidth]{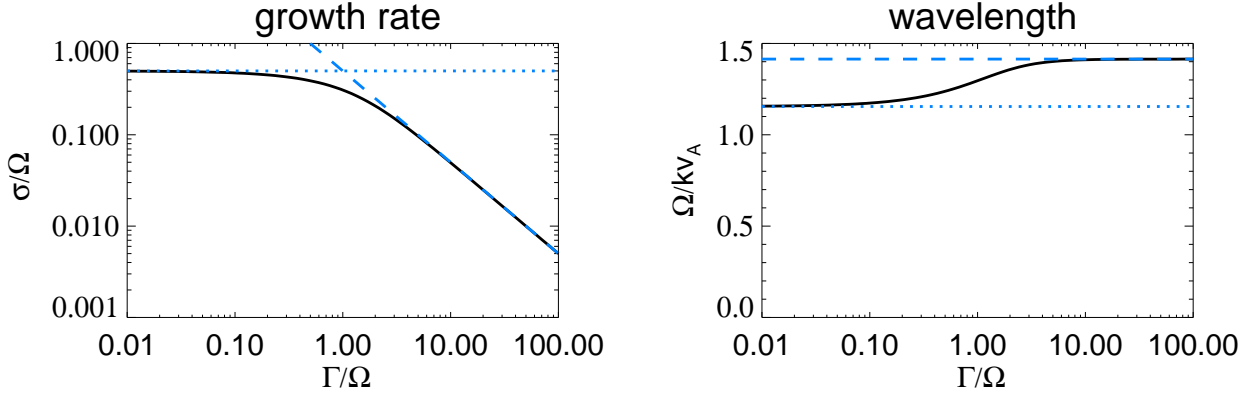}
 \caption{Growth rate (left) and wavelength (right) of the fastest growing MRI mode at scales shorter than the neutrino mean free path as a function of the damping rate $\Gamma$ (assuming zero resistivity). The numerical solution of the dispersion relation (equation~(\ref{eq:dispersion_drag})) is shown with the solid black lines, the strongly damped limit $\Gamma\gg \Omega$ (equations~(\ref{eq:sigma_drag})-(\ref{eq:k_drag})) with dashed lines, and the ideal limit $\Gamma=0$ (Section~\ref{sec:ideal_MRI}) with dotted lines. All quantities are shown in a non-dimensional way: $\Gamma$ and $\sigma$ are normalized by the angular frequency $\Omega$, while $\Omega/(kv_A)$ is a non-dimensional measure of the wavelength.}
             \label{fig:drag_MRI}%
\end{figure*}

Fig.~\ref{fig:drag_MRI} shows the growth rate and wavelength of the fastest growing MRI mode (the numerical solution of the dispersion relation for $\eta=0$ is shown with the solid black line) as a function of the dimensionless parameter $\Gamma/\Omega$ characterizing the effect of neutrino drag on the MRI. The growth of MRI channel modes is not much affected by the neutrino drag as long as $\Gamma<\Omega$. Interestingly, the damping rate threshold $\Gamma \sim \Omega$ is equivalent within a numerical factor to the limit of applicability of the viscous formalism (at which the viscous wavelength equals the neutrino mean free path). This comes from equation~(\ref{eq:k_viscous}) giving the viscous wavelength $\lambda^2 \sim \nu/\kappa $ and the facts that $\Gamma \sim \nu/\langle l_\nu\rangle^2$ (as noted in the previous subsection) and $\kappa \sim \Omega$. As a consequence, in the outer layer of the PNS, where the viscous formalism does not apply, the neutrino drag has little impact on the MRI growth.

When the damping rate $\Gamma$ is increased further, the growth rate of the MRI is reduced significantly while the wavelength of the fastest growing mode changes only slightly\footnote{It might seem surprising at first sight that the wavelength of the most unstable mode varies (although only slightly), while the damping rate is independent of wavelength. For a given magnetic field strength, the modes at different wavelengths have different structures, in particular different ratio of velocity to magnetic field perturbations. Since the damping acts only on the velocity field (but not on the magnetic field) they are affected in a different way by the neutrino drag, even if the damping rate itself does not depend on the wavelength. The situation is different, however, if we change simultaneously the wavelength and the magnetic field strength. Indeed, modes with the same dimensionless wavenumber $kv_A/\Omega$ have the same (dimensionless) structure and are therefore affected by neutrino drag in exactly the same way. As a consequence, the maximum growth rate is independent of the magnetic field strength.}. A useful analytical solution can be obtained in the asymptotic limit $\Gamma \gg \Omega$ (and therefore $\Gamma \gg \sigma$ since $\sigma < \Omega$) and $\eta=0$. In this limit, the dispersion relation reduces to
\begin{equation}
\left(\Gamma\sigma +k^2v_A^2\right)^2 =  k^2v_A^2\left(4\Omega^2- \kappa^2 \right).
\end{equation}
The growth rate of the MRI as a function of vertical wavenumber is therefore 
\begin{equation}
\sigma = \frac{kv_A}{\Gamma}\left(\sqrt{2q}\Omega - kv_A \right),
	\label{eq:sigma_vs_k_drag}
\end{equation}
and the growth rate and wavenumber of the fastest growing mode are
\begin{equation}
\sigma = \frac{q}{2}\frac{\Omega^2}{\Gamma},
	\label{eq:sigma_drag}
\end{equation}
and
\begin{equation}
k = \sqrt{q/2}\frac{\Omega}{v_A},
	\label{eq:k_drag}
\end{equation}
respectively. The asymptotic limits $\Gamma \gg \Omega$ (drag regime) and $\Gamma \ll \Omega$ (i.e. the ideal regime described in Section~\ref{sec:ideal_MRI}) are compared to the full numerical solution in Fig.~\ref{fig:drag_MRI}. An agreement within $10\%$ is obtained for $\Gamma/\Omega <0.2$ (ideal regime) and $\Gamma/\Omega > 3$ (drag regime), and within $20\%$ for $\Gamma/\Omega <0.4$ (ideal regime) and $\Gamma/\Omega > 2$ (drag regime).

The analytical solution shows that compared to the ideal MHD case, the maximum growth rate is reduced by a factor $\Gamma/\Omega$. Importantly, and contrary to the viscous regime, the growth rate remains independent of the magnetic field strength. This comes from the fact that the neutrino drag is independent of the wavenumber, and as a consequence affects the MRI growing on a weak magnetic field (with short wavelength) in the same way as if it were growing on a stronger magnetic field (with longer wavelength). 
As a consequence, if the initial magnetic field is weak the MRI is more likely to grow in this regime than in the viscous regime (where the growth rate is proportional to magnetic field strength). The condition for the MRI to grow at a minimum growth rate $\sigma_{\rm min}$ is independent of the magnetic field strength, and can be cast as an upper limit on the neutrino damping rate 
\begin{equation}
\Gamma < q\Omega^2/(2\sigma_{\rm min}).
	\label{eq:gamma_max}
\end{equation} 

\begin{figure}
\centering
 \includegraphics[width=\columnwidth]{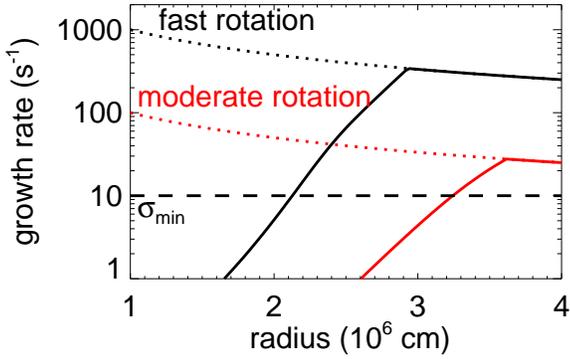}
 \caption{Radial profile of the maximum growth rate of the MRI at scales shorter than the neutrino mean free path. The growth rate (solid lines) is computed by applying equation~(\ref{eq:sigma_drag}) where $\Gamma>\Omega$, and equation~(\ref{eq:sigma_ideal}) where $\Gamma<\Omega$. The dotted lines show the growth rate in the ideal MHD limit (equation~(\ref{eq:sigma_ideal})). The two colours represent the two different normalizations of the rotation profile: fast rotation (black) and moderate rotation (red). Finally, the dashed line shows the minimum growth rate $\sigma_{\rm min}=10\,{\rm s^{-1}}$.}
              \label{fig:sigma_drag}%
\end{figure}

Applying the analytical results of this section to the numerical model of the PNS allows one to compute the growth rate of the MRI as a function of radius in the PNS. The result is shown in Fig.~\ref{fig:sigma_drag} for the two rotation profiles considered. Due to the large variation of the neutrino damping rate inside the PNS, different MRI growth regimes are encountered depending on the radius. Deep inside the PNS, the very large neutrino drag suppresses the growth of the MRI : this is due to the high temperature. Near the PNS surface on the contrary, the neutrino drag does not have much effect on the growth of the MRI (because $\Gamma<\Omega$), which therefore occurs in the ideal regime described in Section~\ref{sec:ideal_MRI}. Finally, at intermediate radii, the neutrino drag has a significant impact on the MRI growth rate but still allows a sufficiently fast growth (we call this the drag regime). The extent of the three different regimes depends on the rotation frequency : the MRI can grow in a larger portion of the PNS for fast rotation than for moderate rotation. Indeed if $\Omega$ is larger the MRI can grow in the ideal regime for higher values of $\Gamma$, and therefore smaller radii. The extent of the drag regime also depends sensitively on $\Omega$: equation~(\ref{eq:sigma_drag}) shows that in this regime the MRI growth rate has a steeper dependence on the rotation rate ($\sigma_{\rm max} \propto \Omega^2$) than in the ideal MHD case ($\sigma_{\rm max} \propto \Omega$). This explains why the region where the MRI grows in the drag regime is less extended in the case of moderate rotation.

\begin{figure}
\centering
 \includegraphics[width=\columnwidth]{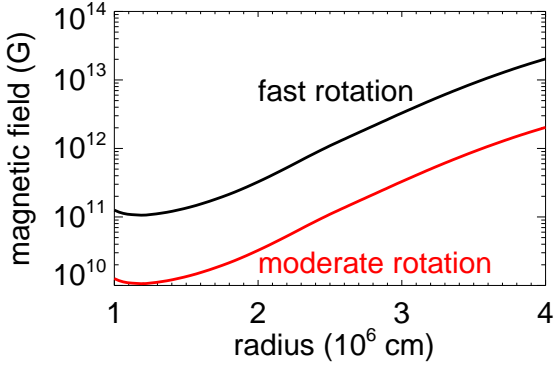}
 \caption{Magnetic field strength at which the wavelength of the fastest growing MRI mode equals the mean free path of electron neutrinos $1/\langle \kappa_\nu \rangle$ (as defined by equation~(\ref{eq:energy_average_drag})): below this critical strength the formalism used in Section~\ref{sec:drag_MRI} is self consistent. The two colours represent the two different normalizations of the rotation profile: fast rotation (black) and moderate rotation (red).}
              \label{fig:lnu_neutrino_drag}%
\end{figure}

Finally, we should determine the condition for the formalism developed in this section to be self-consistent, i.e. that the wavelength of fastest MRI growth be shorter than the neutrino mean free path. As noted above, the wavelength of the fastest growing mode is of the same order of magnitude as in the ideal MHD case, and is therefore proportional to the magnetic field strength. This is again in contrast to the viscous case, where the wavelength of the fastest growing mode is set by the viscosity and angular frequency (independently of the magnetic field strength). The results of this section are self-consistent if the wavelength of the fastest growing MRI mode is shorter than the neutrino mean free path. Using equation~(\ref{eq:k_drag}), this condition can be expressed as an upper limit on the magnetic field strength,
\begin{equation}
B<\sqrt{\rho q/2\pi} \Omega/\langle\kappa_\nu\rangle,
	\label{eq:B_consistency_drag}
\end{equation}
because the relevant neutrino mean free path is the inverse of the opacity averaged over neutrino energy as defined by equation~(\ref{eq:energy_average_drag}). Fig.~\ref{fig:lnu_neutrino_drag} shows this maximum magnetic field strength for consistency of MRI growth in the drag regime. It is quite weak ($10^{10}-10^{11}\,{\rm G}$) deep inside the PNS (where the MRI anyway does not grow efficiently due to the strong drag), and increases by two orders of magnitude towards the surface of the PNS reaching values of $2\times 10^{12}\,{\rm G}$ for moderate rotation and $2\times 10^{13}\,{\rm G}$ for fast rotation. This shows that the growth of the MRI at length-scales shorter than the neutrino mean free path is relevant for weak to moderate initial magnetic fields in the outer parts of the PNS.

%%%%%%%%%%%%%%%%%%%%%%%%%%%%%%%%%%%%%%%%%%%%%%%%%%%%%%%%%%%%%%%%%%%%%%%%%%%%%%%%%%
\section{Discussion and conclusion}
	\label{sec:conclusion}

\begin{figure*}
\centering
 \includegraphics[width=\columnwidth]{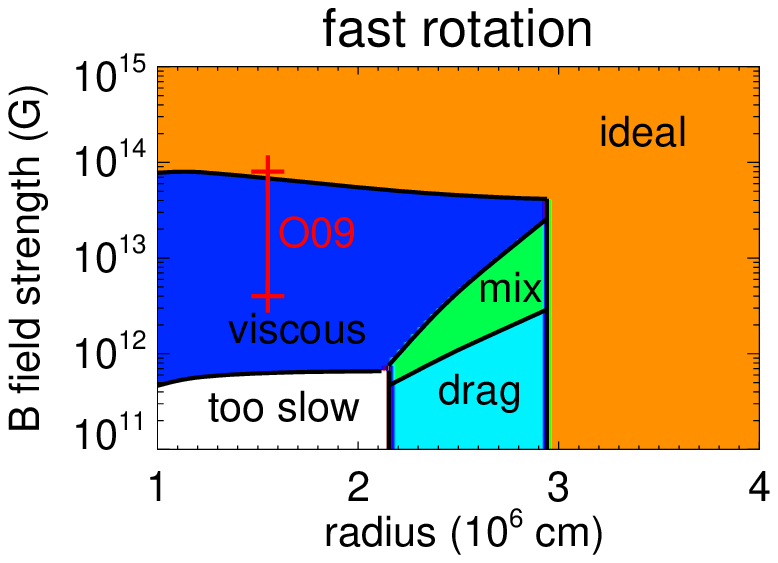}
  \includegraphics[width=\columnwidth]{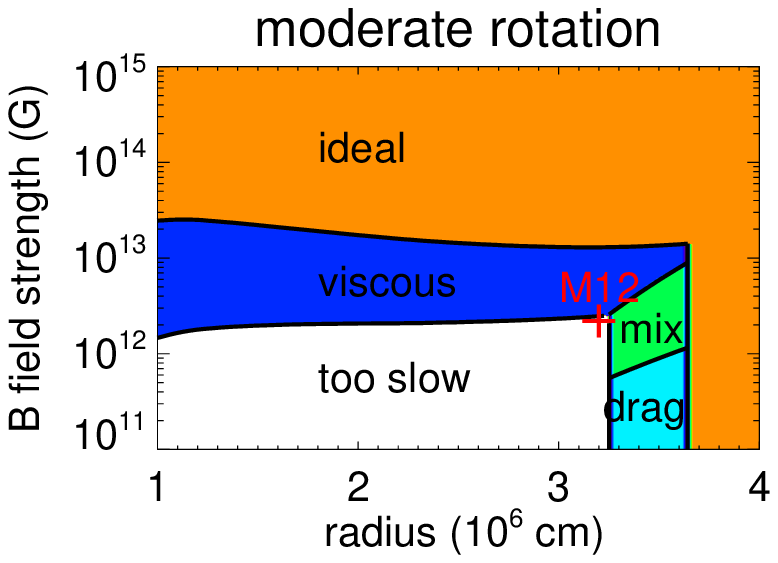}
 \caption{Different regimes of MRI growth as a function of radius and magnetic field strength in the case of fast rotation (left-hand panel) and moderate rotation (right-hand panel), for a PNS model at $t=170\,{\rm ms}$ post-bounce. See the text for a description of the different regimes. The parameter range used in the simulations by \citet{obergaulinger09} is shown in red on the left panel, and the parameters assumed by \citet{masada12} are shown with a red cross on the right panel. The black lines separating the different regimes are defined as follows. The vertical line separating the "too slow" and drag MRI regimes corresponds to equation~(\ref{eq:gamma_max}). The vertical line between the drag and ideal MRI regimes corresponds to $\Gamma=\Omega$ (which also represents within a numerical factor the condition that the viscous MRI wavelength is longer than the neutrino mean free path). The line separating the "too slow" and viscous regimes corresponds to equation~(\ref{eq:viscous_Bmin}). The almost horizontal line separating the viscous and ideal MRI regimes is given by equation~(\ref{eq:Bvisc}). The line separating the drag and mixed regimes shows where the MRI wavelength in the drag regime equals the electron neutrino mean free path (equation~(\ref{eq:B_consistency_drag})). Finally, the line separating the mixed and viscous regimes is defined by the equality of the growth rates in the viscous and the drag regimes (equation~(\ref{eq:B_visc-drag})). The last criterion is only approximate, based on the assumption that the growth rate in the mixed regime lies in between those predicted by the drag and viscous formalisms.}
              \label{fig:MRI_regimes}%
\end{figure*}

In this paper we have studied the impact of neutrino radiation on the growth of the MRI. We have shown that, depending on the physical conditions, the MRI growth can occur in three different regimes:
\begin{itemize}
\item {\bf Ideal regime} (orange colour in Fig.~\ref{fig:MRI_regimes}) : this is the classical MRI regime which applies when neutrino viscosity or drag are unimportant, i.e. if $E_\nu >1$ and $\Gamma/\Omega < 1$. The growth rate of the MRI is then a fraction of the angular frequency independently of the magnetic field strength, while the most unstable wavelength is proportional to the magnetic field strength.
\item {\bf Viscous regime} (dark blue colour in Fig.~\ref{fig:MRI_regimes}) : on length-scales longer than the neutrino mean free path, neutrino viscosity significantly affects the growth of the MRI if $E_\nu<1$. The growth of the MRI is then slower and takes place at longer wavelength compared to the ideal regime. In the viscous regime, the wavelength of the most unstable mode is independent of magnetic field strength, while the growth rate is proportional to the magnetic field strength. As a result, a minimum magnetic field strength of $\sim 10^{12}\,{\rm G}$ is required for the MRI to grow on sufficiently short time-scales.
\item {\bf Drag regime} (light blue colour in Fig.~\ref{fig:MRI_regimes}) : on length-scales shorter than the neutrino mean free path, neutrino radiation exerts a drag on moving fluid elements. This drag has a significant impact on the MRI if the damping rate is larger than the rotation angular frequency ($\Gamma > \Omega$). In this regime, the growth rate of the most unstable mode is independent of the magnetic field strength, but is reduced by a factor $\Gamma/\Omega$ compared to the ideal regime. The wavelength of the most unstable mode is not much affected by the neutrino drag.   
\end{itemize}

\begin{figure*}
\centering
 \includegraphics[width=\columnwidth]{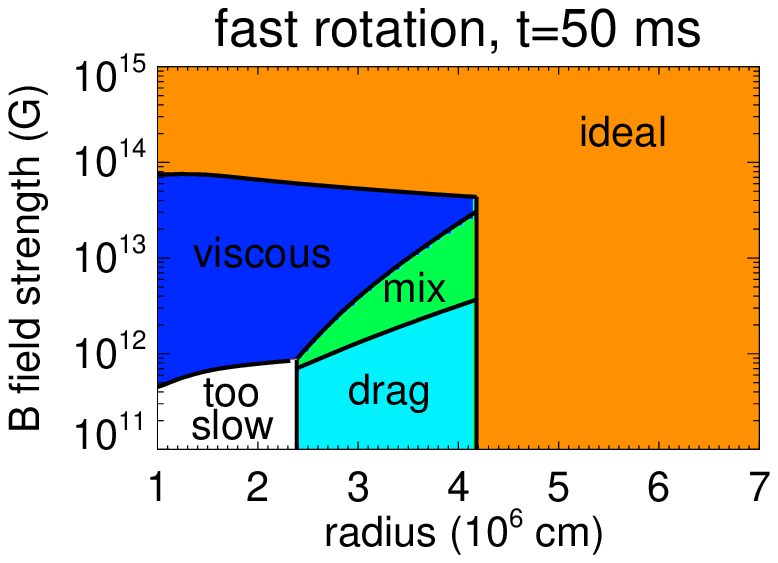}
  \includegraphics[width=\columnwidth]{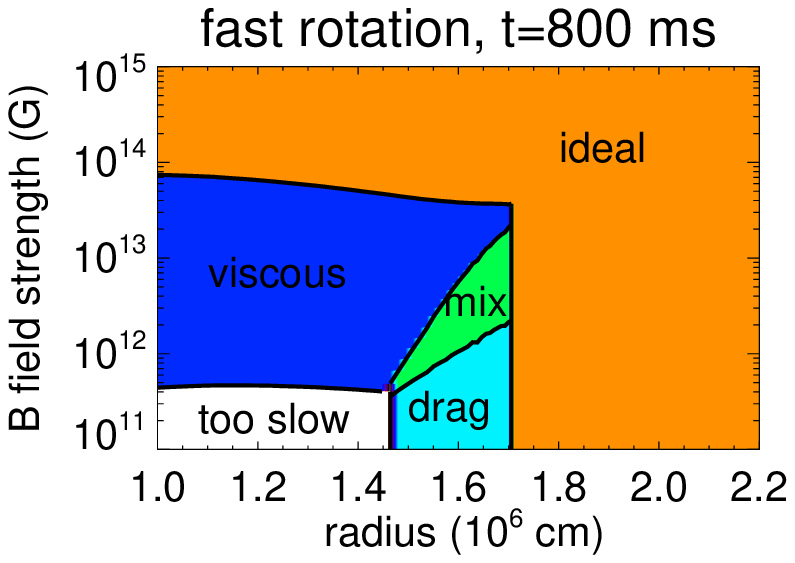}
 \caption{Same as the left panel of Fig.~\ref{fig:MRI_regimes} but at two other times : $50\,{\rm ms}$ after bounce (left-hand panel), and $800\,{\rm ms}$ after bounce (right-hand panel). Note that the horizontal scales are different because they have been rescaled to the size of the PNS.}
              \label{fig:MRI_regimes_different_times}%
\end{figure*}

Fig.~\ref{fig:MRI_regimes} shows where in the parameter space these three regimes apply, as a function of radius and magnetic field strength for the two rotation profiles considered in this paper: fast rotation (left-hand panel) and moderate rotation (right-hand panel). Three regions in the PNS can be distinguished:
\begin{itemize}
\item Deep inside the PNS, the neutrino mean free path is much shorter than the wavelength of the viscous MRI, and $\Gamma \gg \Omega$. In this case, the growth of the MRI at scales shorter than the mean free path is strongly suppressed, and the relevant MRI regime is the viscous MRI described in Section~\ref{sec:viscous_MRI}. The MRI can grow on sufficiently short time-scales if the initial magnetic field is above a critical strength given by equation~(\ref{eq:viscous_Bmin}). Viscous effects become unimportant for strong magnetic fields above $B_{\rm visc}$ given by equation~(\ref{eq:Bvisc}).
\item At intermediate radii, the mean free path of neutrinos is still shorter than the wavelength of the viscous MRI, but $\Gamma$ is not too  large such that the MRI can also grow in the drag regime (i.e. equation~(\ref{eq:gamma_max}) is verified). This is therefore an intermediate case where MRI growth can take place both in the viscous regime at wavelengths longer than the neutrino mean free path, and in the drag regime at length-scales shorter than the mean free path. Since the growth rate in the viscous regime is proportional to the magnetic field strength, the growth is faster in the viscous regime above a critical magnetic field strength, which can be obtained by combining equations~(\ref{eq:sigma_viscous}) and (\ref{eq:sigma_drag})
\begin{equation}
B_{\rm visc-drag} = \sqrt{q\pi\rho\nu\kappa}\frac{\Omega}{\Gamma}.
	\label{eq:B_visc-drag}
\end{equation}
Below $B_{\rm visc-drag}$, the growth is predicted to be faster in the drag regime. However, this regime is self-consistent only if the magnetic field strength is weaker than that given by equation~(\ref{eq:B_consistency_drag}) (the MRI wavelength is then shorter than the mean free path of electron neutrinos). In between these two critical strengths, the MRI growth should actually occur in a mixed regime (shown in green in Fig.~\ref{fig:MRI_regimes}) where electron neutrinos are diffusing and thus induce a viscosity, while the other species are free streaming and exert a drag.
\item Near the PNS surface, the viscous regime is irrelevant because the neutrino mean free path is longer than the wavelength of the MRI. Furthermore, in this region the neutrino drag does not affect much the growth of the MRI because the damping rate is smaller than the angular frequency $\Gamma<\Omega$. As a consequence the MRI growth takes place in the ideal regime without much impact of neutrino radiation.
\end{itemize}

For the two rotation rates considered in this paper, the regimes of MRI growth have somewhat different locations in the parameter space. The region where the MRI cannot grow is more extended for moderate rotation than for fast rotation. The viscous and drag regimes are also less extended. For the ideal regime it is a bit more complicated: it extends to lower magnetic fields deep inside for slower rotation (because the wavelength is longer and the MRI is therefore less affected by viscosity), but weak field growth in the ideal regime can take place deeper inside the PNS for fast rotation (as explained in Section~\ref{sec:drag_MRI}).

The results presented so far correspond to a single time frame at $170\,{\rm ms}$ after bounce. In order to study how the different MRI regimes are affected by the PNS contraction, we have performed the same analysis at two other times (shown in Fig.~\ref{fig:MRI_regimes_different_times} for the case of fast rotation only): $50\,{\rm ms}$ after bounce (at which time the PNS radius\footnote{The PNS radius is defined here as the radius at which the density equals $10^{11}\,{\rm g\,cm^{-3}.}$} is $\sim 70\,{\rm km}$), and $800\,{\rm ms}$ after bounce (the PNS radius has then decreased to $\sim 22\,{\rm km}$). Although the size and structure of the PNS is quite different at these different times, the results are strikingly similar: the location of the different MRI regimes is almost identical once rescaled to the size of the PNS, in particular with very similar values of $B_{\rm min}$ and $B_{\rm visc}$ delimitating the viscous regime.

Note that the physical conditions inside the PNS (neutrino viscosity, neutrino damping rate etc.) have been estimated using a one dimensional numerical simulation of core collapse, which considered a non-rotating progenitor. The moderate rotation is expected to change the PNS structure in a negligible way because the ratio of centrifugal to gravitational forces is less than $10^{-3}$ everywhere in the PNS. Fast rotation, on the other hand, should have a significant influence on the structure of the PNS, as the centrifugal force amounts to $4-8\%$ of the gravitational force in the radius range considered. Rotational support is expected to lead to a more extended PNS and lower temperatures along the equatorial direction \citep{kotake04,ott06}. This would change quantitatively the results presented in this paper: for example the lower temperature would lead to smaller values of the neutrino damping rate, such that the MRI would be less affected by the neutrino drag in the outer envelope of the PNS. The centrifugal force also leads to an oblate PNS, and the MRI growth regime will therefore depend on the angular direction in addition to the radial dependence studied in this paper.

Numerical simulations of the MRI in core collapse supernovae have so far neglected the effects of neutrino radiation on the growth of the MRI \citep{obergaulinger09,masada12,sawai13,sawai14}. Our study shows that this assumption is reasonable for the exponential growth of the MRI only in a limited region of the parameter space, namely in the outer region of the PNS or deeper in the PNS but for quite strong initial magnetic fields ($B\gtrsim 3\times10^{13}-10^{14}\,{\rm G}$ depending on the rotation rate\footnote{Note that this criterion only ensures that the linear growth rate is not much reduced by the viscosity, but the non-linear saturation of the MRI may still be affected by viscosity.}). If the initial magnetic field is not very strong, the growth of the MRI deep inside the PNS is strongly affected by neutrino viscosity. The effect of neutrinos should therefore be taken into account in numerical simulations either by adding a viscous or a drag term (depending on the regime of MRI growth) or by directly computing the neutrino transport and back reaction on the velocity field. In order to properly describe the effects of neutrinos on the MRI (viscosity and drag), a neutrino transport scheme should be multidimensional (i.e. not ray by ray) and should include velocity dependent terms.

\citet{masada12} performed local simulations of the MRI assuming a density $\rho=10^{12} \, {\rm g\,cm^{-3}}$ (this corresponds to a radius of $\sim 32\,{\rm km}$ in our PNS model), an initial magnetic field strength $B = 2.2\times 10^{12}\,{\rm G}$ and an angular frequency $\Omega=100\,{\rm s^{-1}}$. This lies just at the limit between the viscous, mixed, and no-MRI regimes in the moderate rotation profile (their rotation is only slightly faster). We therefore conclude that neutrinos should be taken into account under these conditions, though the prescription to be applied (viscosity or drag) is not clear. \citet{obergaulinger09} considered a box localized at $15.5\,{\rm km}$ from the centre of the PNS, with values at the centre of the box of $\rho=2.5\times 10^{13} \, {\rm g\,cm^{-3}}$, $\Omega=1900\,{\rm s^{-1}}$ (i.e. close to the fast rotation profile we considered), and different values of the magnetic field strength varying between $B = 4\times 10^{12}$ and $8\times 10^{13}\,{\rm G}$. Under these conditions, the MRI is actually in the viscous regime and the neutrino viscosity should therefore be taken into account in the numerical simulation.

The large value of the neutrino viscosity has another important consequence: since the resistivity is quite small in comparison, this leads to a huge value of the magnetic Prandtl number (the ratio of viscosity to resistivity): ${\cal P}_{m} \equiv \nu/\eta \sim 10^{13}$ \citep{thompson93,masada07}. Studies in the context of accretion discs have shown that the level of MRI turbulence is very sensitive to the magnetic Prandtl number, and that it generally increases with this number \citep[e.g.][]{lesur07,fromang07b,longaretti10}. If viscosity is not explicitly taken into account in numerical simulations, numerical dissipation will give rise to a numerical magnetic Prandtl number which may depend on the numerical scheme but which should be of order unity \citep{fromang07a,fromang07b}. This is very far from the true regime, and could lead to underestimate strongly the final magnetic energy and stress. Numerical simulations taking explicitly into account neutrino viscosity will therefore be necessary to assess its influence on MRI saturation.

 Let us now discuss the implications of our findings for the explosion mechanism of core collapse supernovae. We have shown that neutrino viscosity and drag have important consequences for the growth time-scale of the MRI, which need to be taken into account. If the magnetic field is initially weak, the MRI growth can be suppressed by neutrino viscosity (if $B\lesssim 10^{12}\,{\rm G}$ in the PNS) or significantly slowed down (if $B\lesssim 10^{13}-10^{14}\,{\rm G}$). 
The most important finding of this paper is probably that, even if the growth of the MRI from very weak magnetic fields is suppressed in the viscous regime deep inside the PNS, the MRI can grow on wavelengths shorter than the mean free path of neutrinos in the outer parts of the PNS. To have an impact on the explosion, it is probably more important that magnetic field amplification takes place in the outer parts of the PNS rather than in its inner parts. In this respect, our findings confirm that the growth of the MRI can be fast enough to play a role in the explosions of fast rotating progenitors. Which impact it has on the explosion will ultimately depend on the non-linear evolution and saturation of the MRI, which set the efficiency of magnetic field amplification. How this non-linear evolution is affected by the neutrino drag is currently unknown and should be the subject of future numerical studies.

In order to highlight the different regimes of MRI growth in a simple way, we have made a number of simplifying assumptions, which are discussed below. First, we have assumed the magnetic field to be purely poloidal. If the azimuthal magnetic field were much stronger than the poloidal one as obtained by \citet{heger05}, the fastest growing perturbations would be non-axisymmetric \citep{masada06}. In a local analysis (like this article or \citet{masada06}), these shearing waves are only transiently growing typically during a few shear time-scales, due to the fact that their radial wave vector increases linearly with time (and proportionally to the azimuthal wave vector). If their growth is slowed down by the effect of viscosity, it may be difficult for these transiently growing waves to achieve a significant amplification. A WKB analysis as performed by \citet{masada07} is not valid in that case (these authors assumed the azimuthal magnetic field to be strong enough such that viscous effects do not reduce the growth significantly), but the transient amplification may instead be meaningfully studied using a non-modal approach \citep{squire14a,squire14b}. It would also be interesting to study the non-axisymmetric growth in the presence of neutrino drag. 

Secondly, we have neglected buoyancy effects, which can arise due to the presence of entropy and composition gradients. Buoyancy can be either stabilizing or destabilizing depending on the location in the PNS. A convective region (i.e. destabilizing buoyancy) is thought to be present in the inner part of the PNS, at radii between $10-15\,{\rm km}$ and $20-30\,{\rm km}$ \citep[and references therein]{keil96,dessart06,buras06b}. The convective motions could play an important role in amplifying the magnetic field, in particular in the region of parameter space where the MRI growth is too slow. Outside of this convective region, buoyancy has a stabilizing effect and can potentially have a strong impact on the MRI growth \citep{balbus94,menou04,masada06,masada07,obergaulinger09}. In the neutrino diffusive regime, thermal and lepton number diffusion can, however, alleviate the stabilizing effect of buoyancy on the MRI \citep{acheson78,menou04,masada07}. This was demonstrated in the context of a PNS by \citet{masada07}, who considered the case of an azimuthal magnetic field that is much stronger than the vertical component, and also strong enough so that the (non-axisymmetric) MRI is not much affected by the neutrino viscosity (i.e. $B>B_{\rm visc}$ in our notations). Under these conditions, they showed that the MRI can grow at a rate close to its maximum growth rate despite the stabilizing effect of buoyancy, if the thermal and lepton number diffusivities are much larger than the viscosity (more precisely if $\chi \gtrsim \nu N/\Omega$, where $\chi$ represents either thermal or lepton number diffusivity, and $N$ is the Brunt-V\"ais\"al\"a frequency). However, they did not much explore the regime of less strong magnetic fields, where the MRI is strongly affected by neutrino viscosity. It would be very instructive to investigate the effect of buoyancy in this viscous regime by applying the formalism of \citet{masada07} to the PNS model considered here, but this is beyond the scope of the present work and is therefore left for a future study. Finally, we note that buoyancy effects on the MRI growth at wavelengths shorter than the neutrino mean free path are so far unknown. They should be studied in the future, since they are probably relevant in the outer parts of the PNS.

\section*{Acknowledgements}
We thank Florian Hanke for providing us with his simulation data. We thank the anonymous referee for his/her careful report, which helped us improve the presentation of the paper. JG acknowledges support from the Max-Planck--Princeton Center for Plasma Physics.

\bibliography{supernovae}

\bsp
\label{lastpage}

\end{document}